\documentclass{PoS}
\graphicspath{ {images/} }

\title{Survey of nearby active galactic nuclei with the HAWC Observatory}
\ShortTitle{AGN Survey with HAWC}

\author{\speaker{Zhixiang Ren}\thanks{Presented by Sara Couti\~{n}o de Le\'{o}n}\\
        The University of New Mexico\\
        E-mail: \email{zxren@unm.edu}}
        
\author{Robert Lauer\\
        The University of New Mexico\\
        E-mail: \email{rlauer@phys.unm.edu}}

\author{John A.J. Matthews\\
        The University of New Mexico\\
        E-mail: \email{johnm@phys.unm.edu}}

\author{For the HAWC Collaboration\\
        For a complete author list, see http://www.hawc-observatory.org/collaboration/}
        
\abstract{
The High Altitude Water Cherenkov Observatory (HAWC) has a wide field-of-view (FOV, $\sim$2sr) and a high duty cycle ($\sim$95\%), which make it a powerful survey and monitoring experiment for sources of TeV gamma rays. We present a systematic survey of gamma-ray sources based on the Fermi 3FHL catalog. Sources are restricted to HAWC's FOV (Declination 19$^\circ$ $\pm$ 40$^\circ$) and to extragalactic sources with redshift: 0.001 $\textless$ z $\textless$ 0.3. Extragalactic gamma-ray sources are dominated by active galactic nuclei (AGN) and TeV gamma-ray sources are mostly BL Lac-type blazars. The study of AGNs through high energy gamma rays has opened a new window into the extreme processes of particle acceleration in the jets of these objects and provides a way to study the photon propagation and extra-galactic background light. We have improved the HAWC sensitivity at low energies (100 GeV to 1 TeV) based on the Crab pulsar, which is an excellent calibration source for TeV gamma rays. We will present the results of searching for and monitoring nearby AGNs with the improved analysis. 
}

\FullConference{35th International Cosmic Ray Conference --- ICRC2017\\
		10--20 July, 2017\\
		Bexco, Busan, Korea}

\begin{document}

\section{Introduction}
\begin{figure}[h]
    \centering
    \includegraphics[width=6.0in]{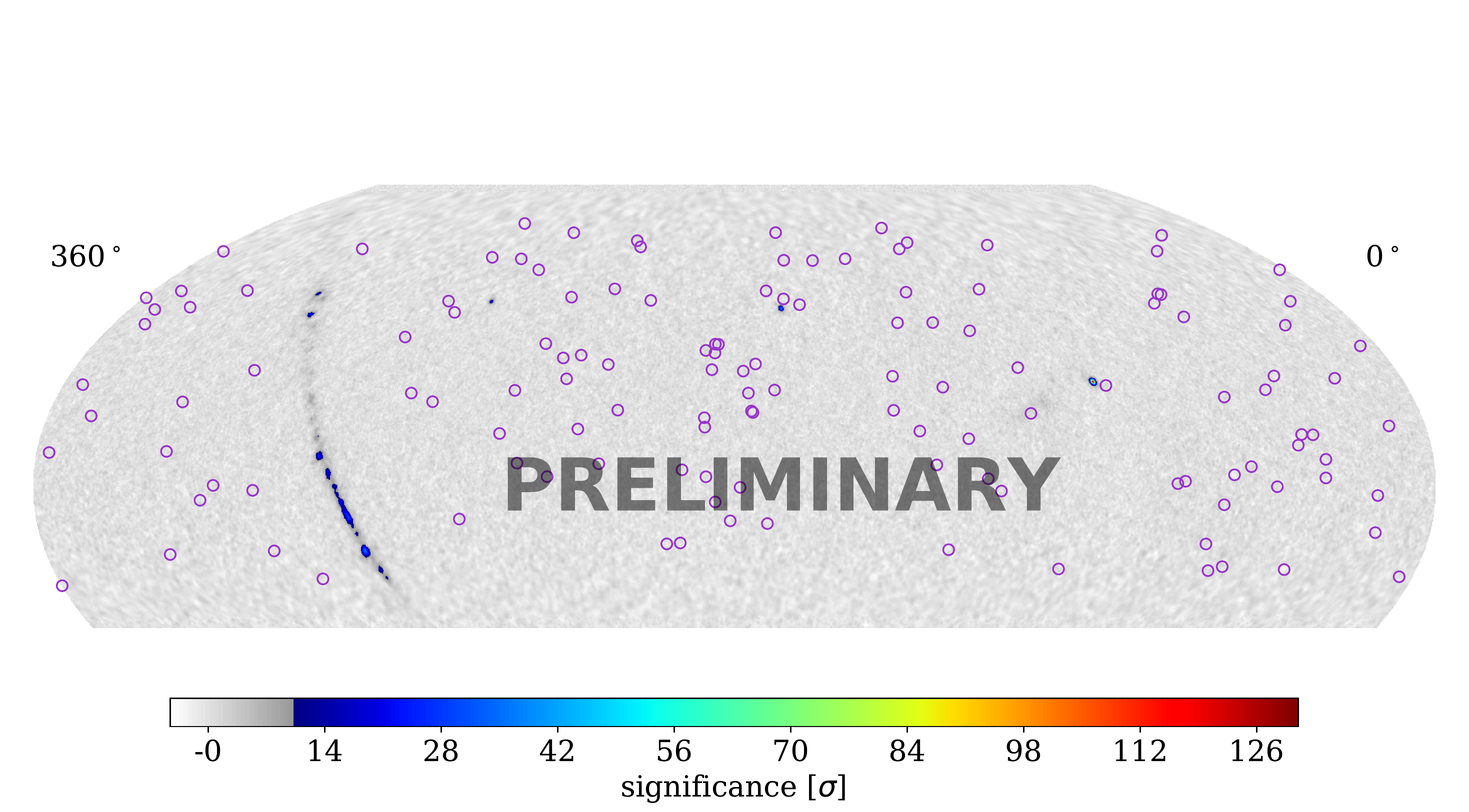}
    \caption{The position of selected 132 sources (excluding Mrk421 and Mrk501) for this survey. The figure is in equatorial coordinates and each circle represents one source. The points with high significances are sources detected by HAWC \cite{catalog_paper}.}
    \label{fig:source_position}
\end{figure}

Active galactic nuclei (AGN) are the most numerous class of identified very-high-energy (VHE) $\gamma$-ray sources \footnote{TeVCat: http://tevcat.uchicago.edu/}. So far, HAWC has detected two of the brightest VHE AGNs, Markarian 421 (Mrk 421) and Markarian 501 (Mrk 501) \cite{lightcurve_paper}. In this contribution, we present a systematic survey of AGN locations with the combined data from over two years of observations from the HAWC Observatory. With a duty cycle of $\sim$95$\%$ and a daily exposure of $\sim$6 hours for any location in 2/3 of the sky, HAWC provides unique long-term monitoring data that makes it possible to extend surveys like those of Fermi-LAT to TeV energies. We select sources based on HAWC's FOV (Declination 19$^\circ$ $\pm$ 40$^\circ$) and redshift limits (0.001 $\textless$ z $\textless$ 0.3) for extragalactic sources from "The Third Catalog of Hard Fermi-LAT Sources" (3FHL) \cite{3FHL}. To avoid source confusion, we exclude sources within $\pm$5$^\circ$ of the galactic plane and sources closer than 3$^\circ$ to sources detected by HAWC \cite{catalog_paper}. Due to the "$\gamma$-$\gamma$" interaction between the VHE photons and the extragalactic background light (EBL), we do not expect HAWC to detect very far away sources. Therefore we have limits for redshift. We found 134 AGNs identified in the 3FHL that meet these criteria. This source list contains mostly BL Lac objects, as well as a few FSRQs and radio galaxies (e.g. M 87). The survey is based on more than two years data (760 days) from HAWC with a low energy sensitivity improved analysis. Figure \ref{fig:source_position} shows a sky-map with all the selected sources positions. 

\section{Analysis}
\subsection{HAWC Analysis}
The HAWC analysis \cite{crab_paper} uses data bins based on the fraction of available photomultiplier tubes (PMTs) that record light during the particle shower \cite{crab_paper}. Each analysis bin has a set of gamma-hadron (G/H) separation cuts that are based on optimizing Crab significance from HAWC data. In the standard HAWC analysis, the data bin with the smallest particle showers is not included. For this improved analysis, we find a set of G/H separation cuts for the lowest data bin and also re-optimized the G/H separation cuts for the second lowest data bin based on over two years data. In the standard analysis, the Crab nebula is not detected with the lowest data bin with two-year HAWC data, while we have a detection with $\sim$9$\sigma$ for one year data in the newly optimized lowest data bin. The improved analysis enhances HAWC sensitivity to gamma rays in the two lowest data bins, which includes the smallest events that trigger 4.4$\%$ to 10.5$\%$ of $\sim$1200 PMTs. The mean energy of events in the lowest data bin is $\sim$300\,GeV. This is smaller than that from the standard analysis, which is $\sim$1.1\,TeV. For hard spectra, both analyses give very consistent results on upper limits, but the improved analysis has a better performance for small, lower energy showers that are relevant in steep spectra. 

\subsection{Sensitivity}
\begin{figure}[h]
    \centering
    \includegraphics[width=2.95in]{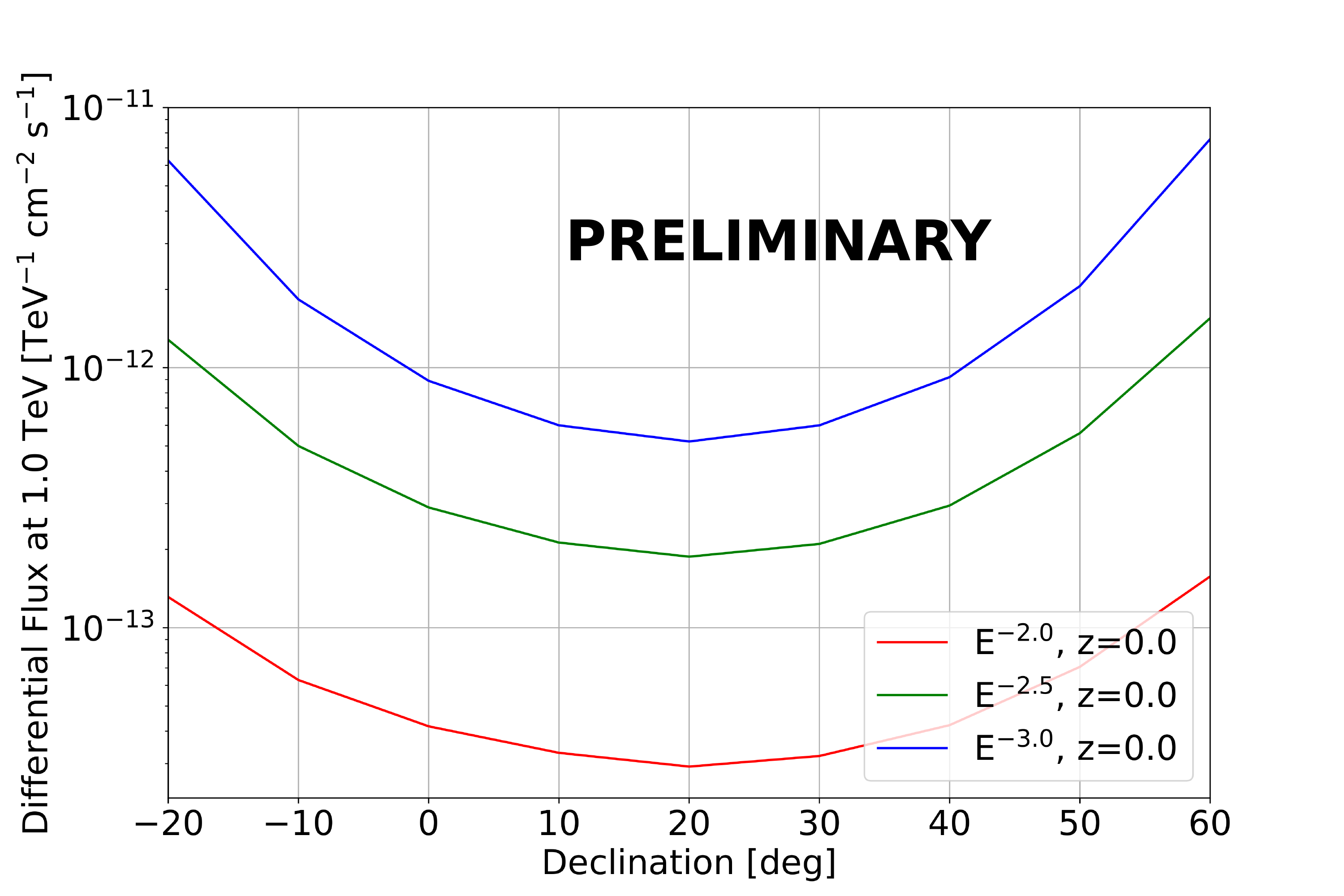}   
    \includegraphics[width=2.95in]{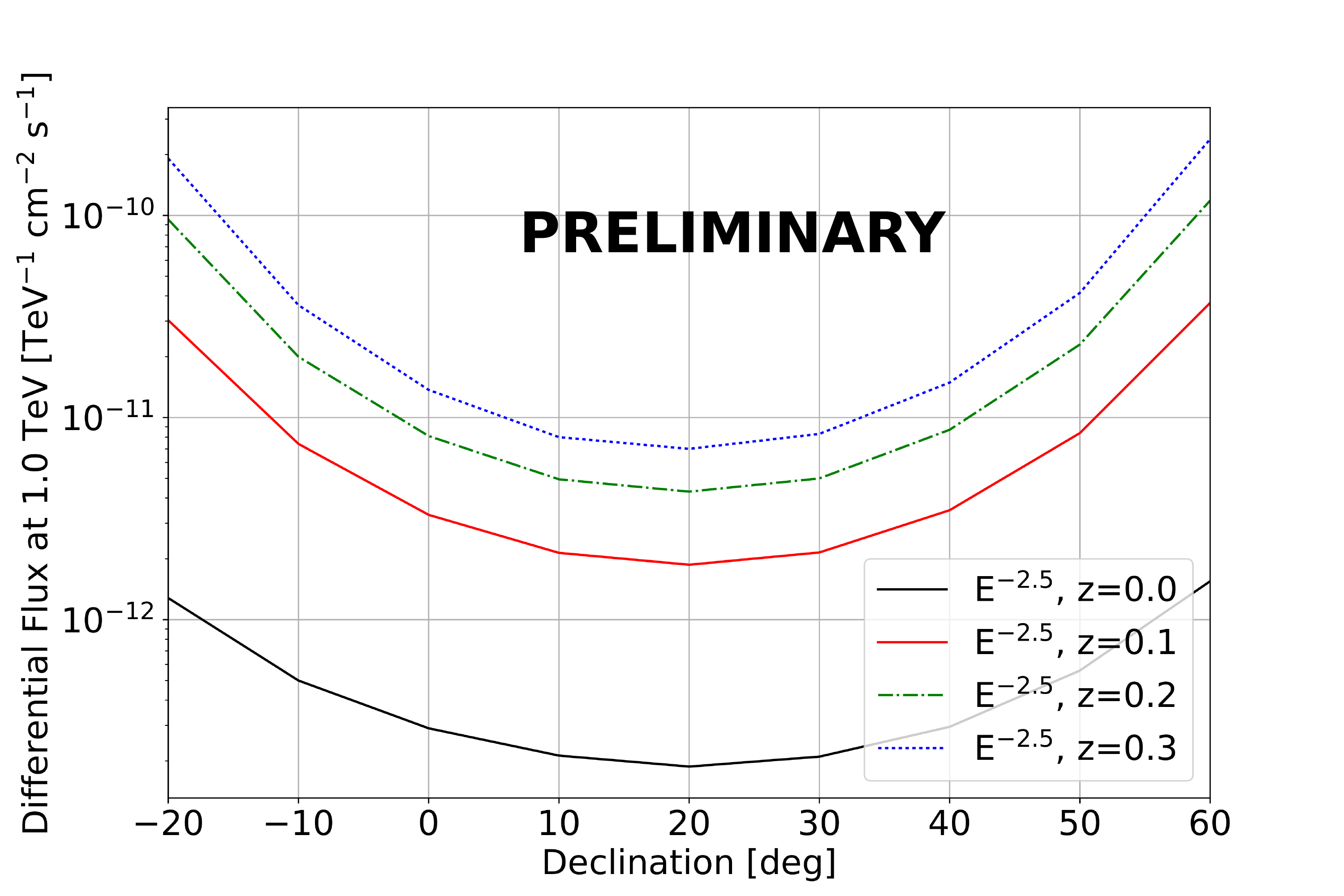}   
    \caption{Sensitivity of HAWC based on 2 years data, where the sensitivity is defined as the 95$\%$ confidence level upper limit. The left figure shows sensitivities for spectra with different power law indices without EBL absorption. The right figure shows the sensitivities to intrinsic spectra with spectrum index -2.5 and different redshifts, including EBL absorption effects.}
    \label{fig:sensitivity_stuff}
\end{figure}

Figure \ref{fig:sensitivity_stuff} summarizes the sensitivities of the improved analysis for different spectrum assumptions. The sensitivity is defined as the 95$\%$ confidence level (95CL) average upper limit from the background-only hypothesis. We employ a likelihood analysis \cite{hawc_likelihood_analysis} that is based on a spectral model to interpret the data. For each source, the intrinsic model is attenuated by the EBL absorption based on the model from \cite{gilmore_2012}. 

\begin{figure}[h]
    \centering
    \includegraphics[width=2.95in]{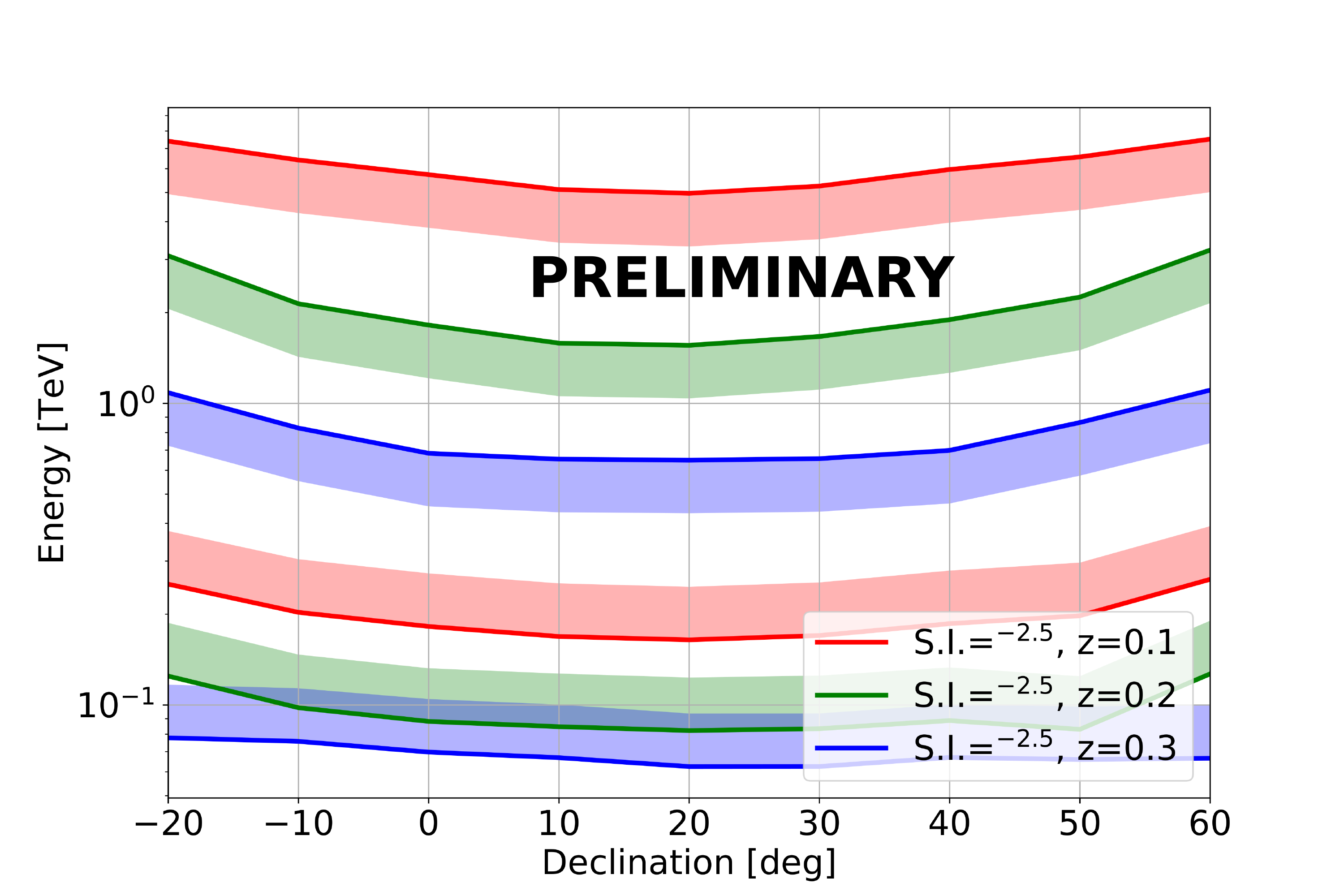}   
    \includegraphics[width=2.95in]{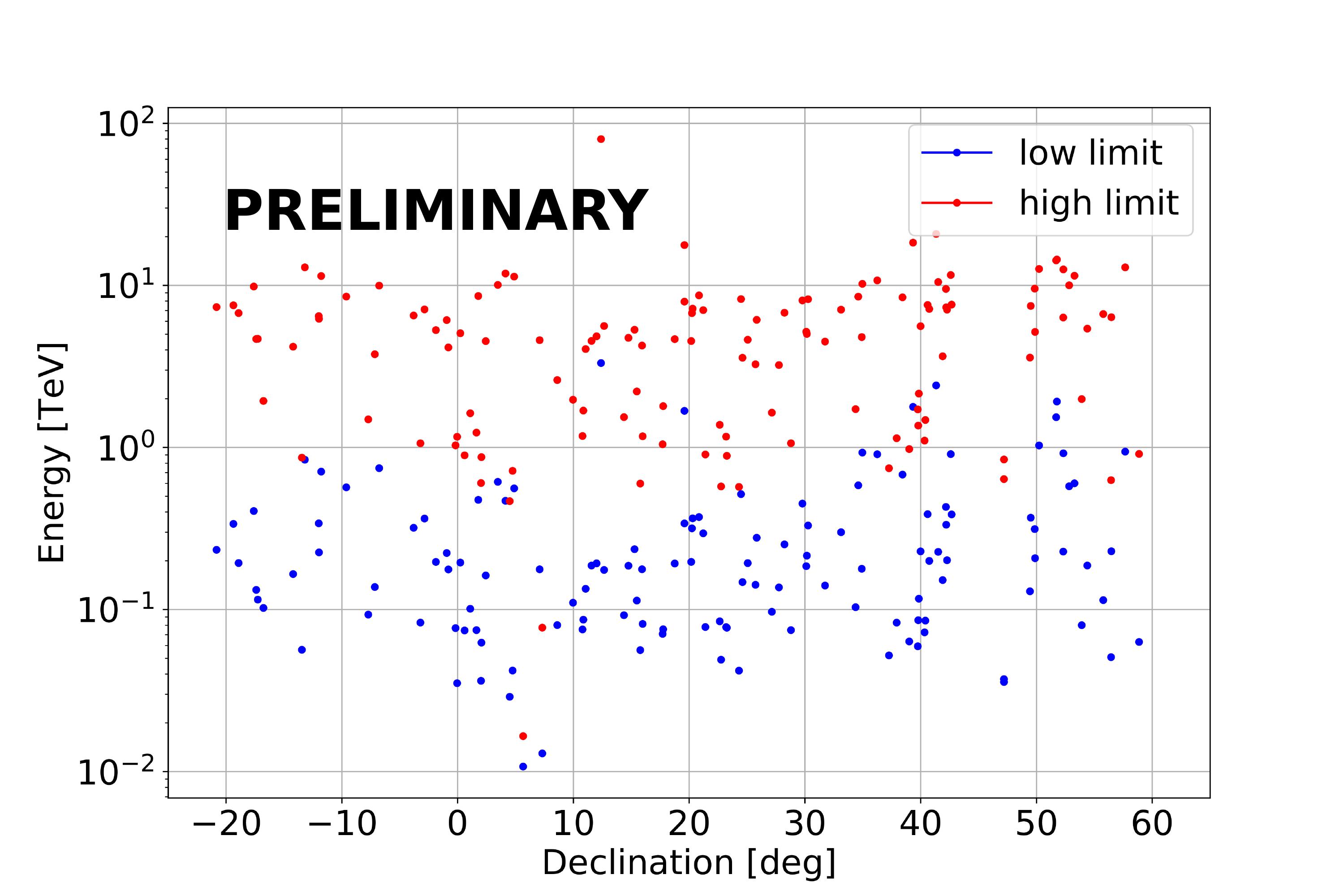}    
    \caption{Left: upper and lower ends of the energy range contributing to the central 75$\%$ of the test statistic of the point source search. Right: the energy ranges of 134 selected 3FHL sources.}
    \label{fig:energy_stuff}
\end{figure}

The left part of Figure \ref{fig:energy_stuff} shows the energy range that contributes 75$\%$ to the test statistic in the point source search \cite{catalog_paper}. The photon energy ranges that are covered by our observations depend on the spectral model and are also calculated for a power law index -2.5 with different redshifts for different declination. The right panel in Figure \ref{fig:energy_stuff} shows the energy range for the different model spectra of the selected 134 3FHL sources.

\section{Results}
\begin{figure}[h]
    \centering
    \includegraphics[width=4.5in]{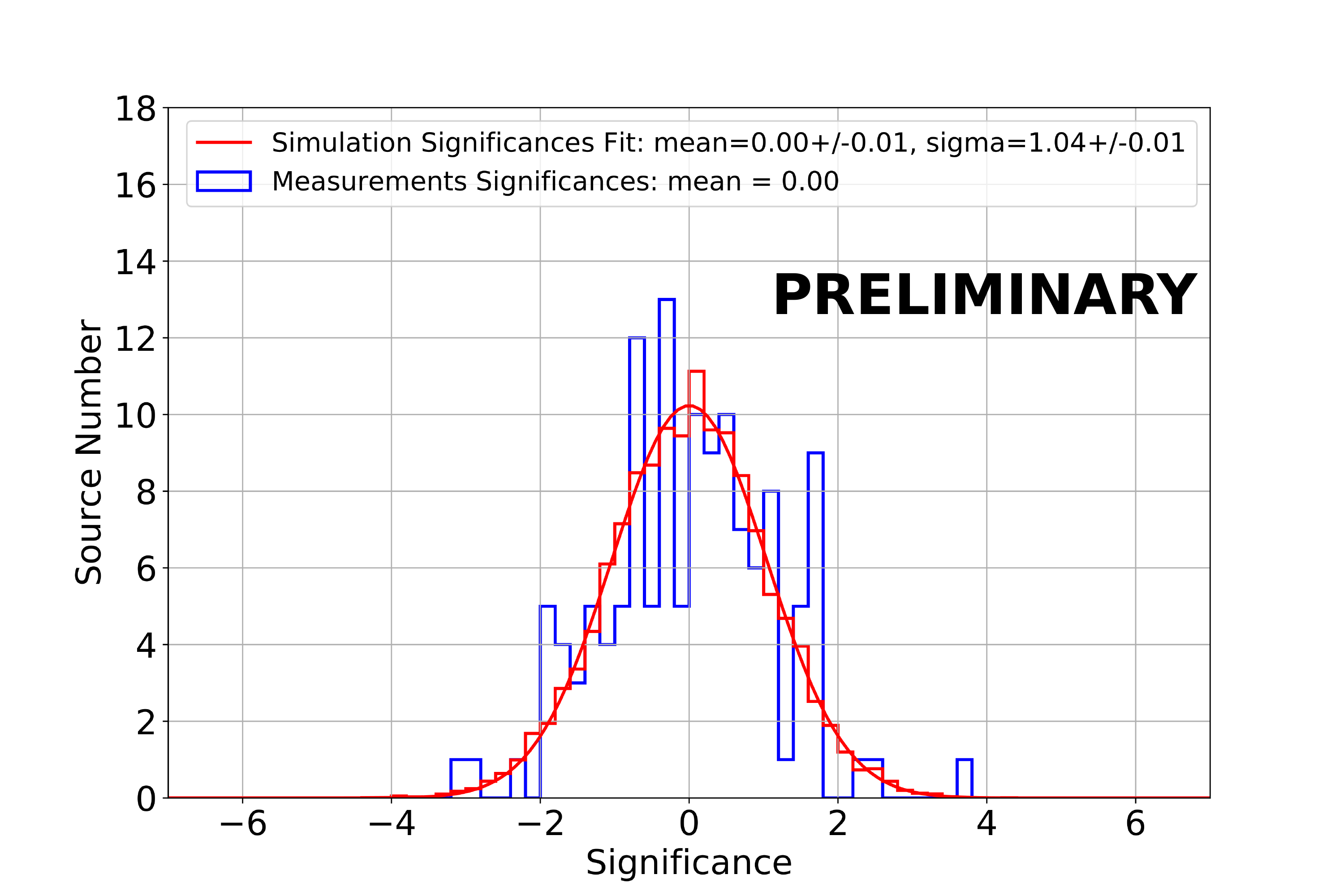}     
    \caption{The significance distribution (in blue) of selected 3FHL sources (excluding Mrk 421 and Mrk 501). The histogram in red (with a Gaussian function fit) shows the significance distribution of random points in the sky.}
    \label{fig:significance_stuff}
\end{figure}

We have applied the analysis to all selected 134 sources and observed significances $\textgreater$ 5$\sigma$ (pre-trial) only for Mrk 421 (39.47$\sigma$) and Mrk 501 (23.17$\sigma$). Therefore, we calculated the 95$\%$ confidence level upper limit for each source excluding Mrk 421 and Mrk 501. The significance distribution of the remaining 132 sources (excluding Mrk 421 and Mrk 501) is shown in Figure \ref{fig:significance_stuff}. The distribution is not skewed to positive values therefore it shows no sign of a statistical detection. To understand the significances distribution in Figure \ref{fig:significance_stuff}, we checked 132 random points in the sky and calculated the significances. We repeated this process for 500 times, normalized the distribution and did a Gaussian function fit, which is also shown in red in Figure \ref{fig:significance_stuff}. The outlier on the right side is VER J0521+211 ($\sim$3.8$\sigma$), which is 3.07$^\circ$ away from the Crab Nebula.

\begin{figure}[h]
    \centering
    \includegraphics[width=5.0in]{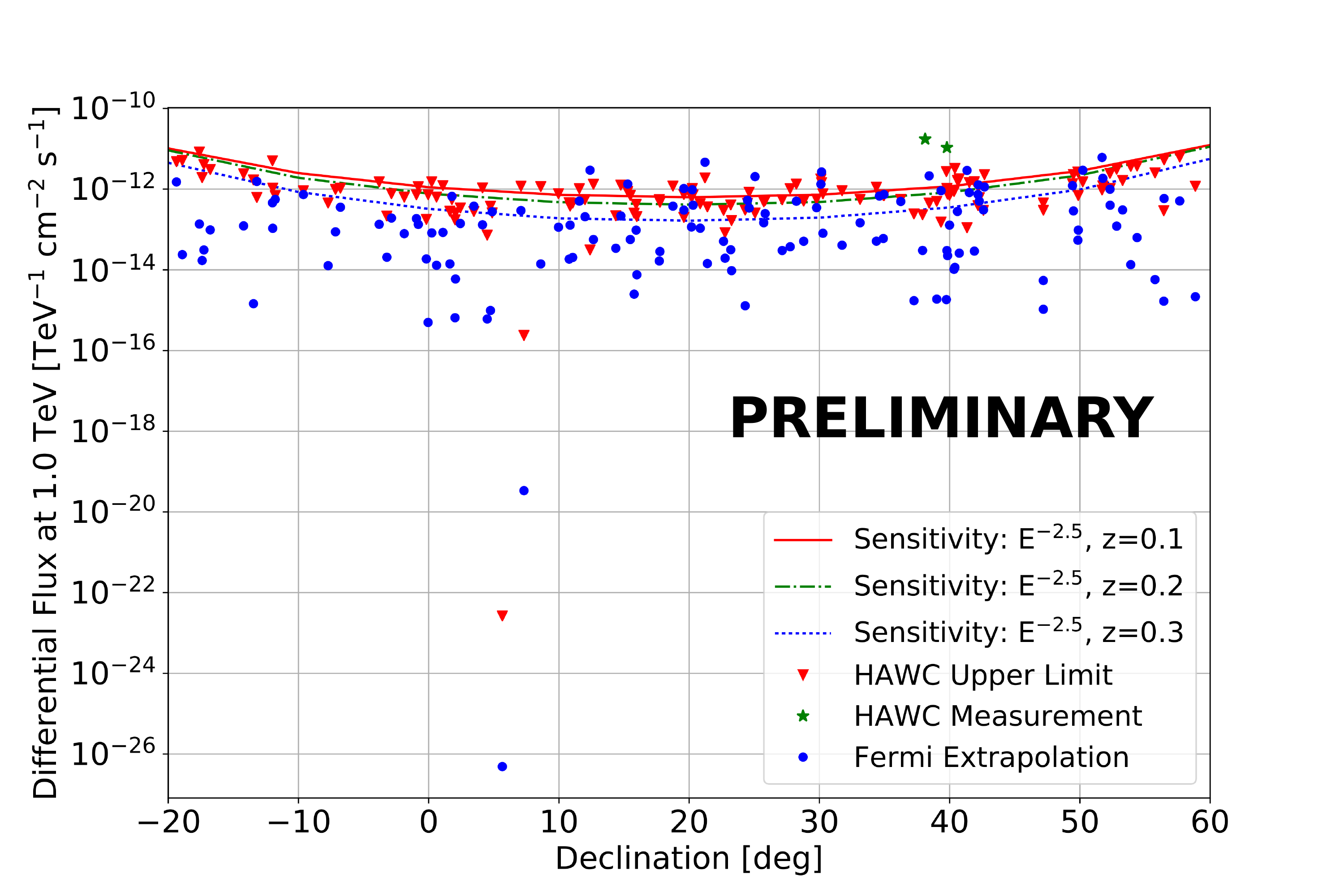}   
    \caption{The HAWC 95CL upper limits (differential flux at 1TeV) vs Fermi-LAT results extrapolations. The red points are the HAWC 95CL upper limits and the blue points are the Fermi-LAT extrapolation results. The HAWC measurements are for two Markarians (green stars). For most sources, the HAWC upper limits are close to the average upper limits. The two sources with very low fluxes are due to the very soft spectra and large redshift.}
    \label{fig:upperlimits_extrapolations}
\end{figure}

Due to the non-detection of the 132 sources, we calculated individual flux limits with the method discussed in \cite{darkmatter_paper} and compared to a direct extrapolation of the Fermi-LAT flux measurement presented in the 3FHL. For both limits and flux extrapolations, we take into account the information from the 3FHL catalog for each individual source including location, redshift and power law spectrum index. The EBL absorption is applied for the calculation of fluxes. Overview of the comparison of HAWC 95CL upper limits and extrapolations from Fermi-LAT results are shown in Figure \ref{fig:upperlimits_extrapolations}. For most source, the HAWC upper limit is higher than the extrapolation result. However there are a total of 23 sources for which HAWC
can constrain the fluxes to lower levels than the extrapolations based on the information from 3FHL. We will present more investigations about these sources in the future.

\begin{figure}[h]
    \centering
    \includegraphics[width=4.0in]{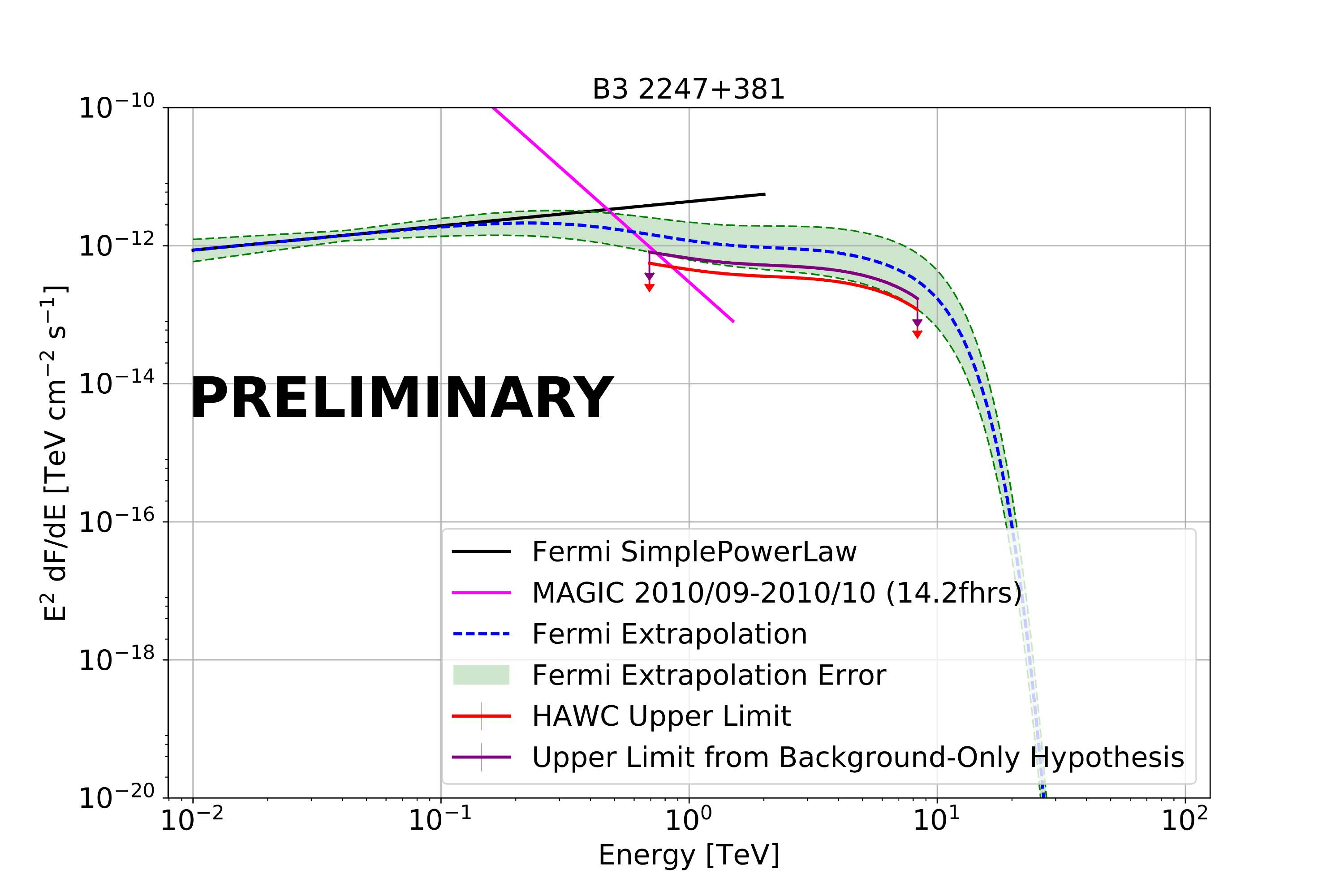}   
    \caption{One example (B3 2247+381) showing how HAWC's 95CL upper limit can constrain the flux to a level lower than Fermi-LAT extrapolation result, based on 3FHL information. We include a short time duration measurement from MAGIC in earlier time. Please see text for more discussion.}
    \label{fig:spectrum_B3}
\end{figure}

\begin{figure}[h]
    \centering  
    \includegraphics[width=4.0in]{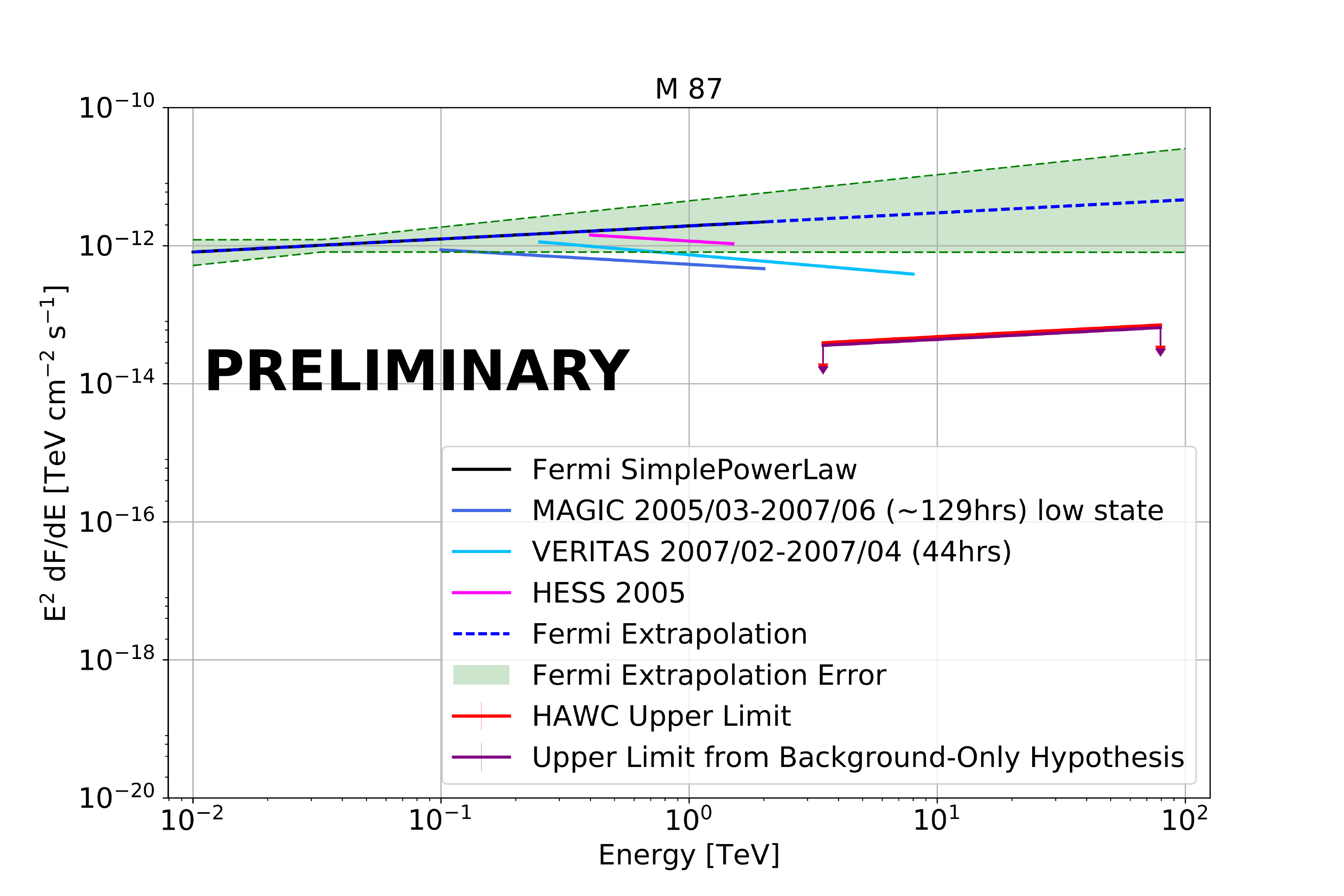}
    \caption{Another example (M 87) showing how HAWC's 95CL upper limit can constrain the flux to a level lower than Fermi-LAT extrapolation result, based on 3FHL information. We include the short time duration measurements from IACTs, which have much softer spectra shape. Please see text for more discussion.}
    \label{fig:spectrum_M87}
\end{figure}

Figure \ref{fig:spectrum_B3} and \ref{fig:spectrum_M87} show spectra of two example sources with constrains from HAWC. The purple curves show the 95CL upper limit from background-only hypothesis. For B3 2247+381 (Figure \ref{fig:spectrum_B3}), the HAWC 95CL upper limit is lower than the Fermi-LAT extrapolation result. The magenta line shows a short time duration measurement from MAGIC  \cite{b32247_magic} before the period covered by HAWC data. The spectral shape from the MAGIC measurement is different than the Fermi-LAT extrapolation. If this spectral shape by MAGIC is a good model for the whole 2-year period that HAWC covers, the non-detection and the HAWC upper limit are consistent with the MAGIC result.

For M 87 (Figure \ref{fig:spectrum_M87}), the HAWC 95CL upper limit is much lower than the Fermi-LAT extrapolation result. We also include short duration measurements from MAGIC \cite{m87_magic}, VERITAS \cite{m87_veritas} and HESS \cite{m87_hess}, all before the HAWC data period. These Imaging Air Cherenkov Telescopes (IACTs) measured softer spectra ($\sim$2.2) from $\textgreater$100GeV to a few TeV, compared to $\sim$1.8 as from 3FHL. The upper limit from HAWC is significantly lower than the flux expected from a simple extrapolation of these IACT's results at higher energies. There may be several explanations for this. The hard spectrum index we used in the context of performing a systematic survey based on the 3FHL information may not be appropriate at multi TeV energies. Furthermore, the flux of M 87 has been observed to be variable and the 2 years that HAWC covers might be dominated by lower flux state compared to the earlier measurements from these IACTs. For a separate discussion of M 87 with HAWC data, see \cite{Daniel_ICRC_2017}.

\section{Summary}
HAWC data from over two years of operation have made possible the first systematic survey of nearby AGNs at TeV energies. Our study is based on sources selected from the Fermi-LAT 3FHL catalog. The selected 134 AGNs include mostly BL Lac objects and some FSRQs and radio galaxies. Out of this list, only Mrk 421 and Mrk 501 have been detected with HAWC, as previously reported. We calculated 95$\%$ confidence level upper limits for the other 132 sources. There are a total of 23 sources for which HAWC can constrain the fluxes below the Fermi-LAT extrapolated results. One of the most constraining upper limits, in view of Fermi-LAT extrapolation and IACT results, is that for M 87. We will perform further investigations of these results and report an overview of all flux limits in an upcoming publication. 

The 3FHL catalog provides the most complete information on gamma-ray emissions so far for AGNs not yet observed at TeV energies. In part to compensate for the difference between the Fermi-LAT energies and HAWC energies, this analysis included the EBL absorption correction. 3FHL sources that have been observed by IACT's, like the examples in Figure \ref{fig:spectrum_B3} and \ref{fig:spectrum_M87}, have potentially been observed in flaring states. Thus, these IACT flux measurements may not represent a long term AGN behavior. The results from HAWC, on the other hand, are the most 
constraining limits on two year average flux averages for these AGNs at TeV energies. Separate analyses of AGNs using different information are the topics of other HAWC studies \cite{Daniel_ICRC_2017}.

\section*{Acknowledgment}
We acknowledge the support from: the US National Science Foundation (NSF); the US Department of Energy Office of High-Energy Physics; the Laboratory Directed Research and Development (LDRD) program of Los Alamos National Laboratory; Consejo Nacional de Ciencia y Tecnolog\'{\i}a (CONACyT), M{\'e}xico (grants 271051, 232656, 260378, 179588, 239762, 254964, 271737, 258865, 243290, 132197), Laboratorio Nacional HAWC de rayos gamma; L'OREAL Fellowship for Women in Science 2014; Red HAWC, M{\'e}xico; DGAPA-UNAM (grants IG100317,
IN111315, IN111716-3, IA102715, 109916, IA102917); VIEP-BUAP; PIFI 2012, 2013, PROFOCIE 2014, 2015;the University of Wisconsin Alumni Research Foundation; the Institute of Geophysics, Planetary Physics, and Signatures at Los Alamos National Laboratory; Polish Science Centre grant DEC-2014/13/B/ST9/945; Coordinaci{\'o}n de la Investigaci{\'o}n Cient\'{\i}fica de la Universidad Michoacana. Thanks to Luciano D\'{\i}az and Eduardo Murrieta for technical support.

\bibliographystyle{JHEP}
\bibliography{references_2017} 
%\begin{thebibliography}{99}
%\bibitem{...}
%....
%\end{thebibliography}

\end{document}